# Nuclear and Radiological Security in a Global Context:

Nick Jones
Bournemouth and Poole College
Dorset BH1 3JJ
jonesn@bpc.ac.uk

**Abstract:** *This paper considers the state of nuclear and radiological security in the UK and abroad and reports on the methods that could be employed by terrorists with radiological or nuclear material to cause destruction. It is shown that despite current safeguards that problems arise due to materials that are unaccounted for and poor implementation of detection regimes in some geographical regions. The prospect of a future terrorist event that involves nuclear or radiological materials seems likely despite best efforts of prevention.*

### 1.0: Introduction

"Mass-destructive terrorism is now the greatest non-traditional threat to international security, and of these, nuclear terrorism poses a real danger. A terrorist use of a nuclear-yield device is the most devastating type of terrorism conceivable and a radiological attack is the most likely variety of non-conventional terrorism." Gavin Cameron, (1999), Nuclear Terrorism: A Threat Assessment for the 21$^{st}$ Century. [1]. Two years after Cameron's book was published an event occurred that made the subject of nuclear terrorism more than academic supposition.

*"The tragic terrorists' attacks on the United States were a wake up call to us all. We cannot be complacent. We have to and will increase our efforts on all fronts – from combating illicit trafficking to ensuring the protection of nuclear materials – from nuclear instillation design to withstand attacks to improving how we respond to nuclear emergencies."*
IAEA Director General Mohamed ElBaradei, 21$^{st}$ September, 2002. [2].

The purpose of this paper is to evaluate nuclear security across spatial boundaries and to assess nuclear and radiological threats in a global context. Events of magnitude highlight vulnerabilities and threats creating the need to reappraise models of protection and the efficacy of safeguards. Terrorism has been an activity on the geopolitical landscape for centuries. Yet the Al-Quaedia attack on the World Trade Centre on 11$^{th}$ September 2001 was by far the most audacious and severe act of terrorism to have been inflicted on the contemporary Western world.

The scale of impact that terrorist acts produce may increase because the greater the destruction/disruption, the more media coverage and the greater the attention drawn to the ideological or political view of the terrorists. Biological or chemical weapons would not convey the same prestige as a nuclear attack. [1].

Barnaby (2003) contends that there are 5 types of terrorist:
- Individual terrorist (e.g. Theodore Kaczynski [The Unabomber])[1].
- Nationalist terrorist (e.g. IRA).
- Political terrorist (e.g. fascists)
- Single-issue terrorist (e.g. animal rights)
- Religious fundamental terrorist (e.g. Al-Quaedia). [3].

Greater freedom of movement internationally (with the exception of the US) has as never before provided the conditions whereby states and sub-state actors intent on malicious acts have the mobility to implement such operations from within the confines of the victim state.

### 2.0: Weapons of Mass Disruption and Destruction:

The following will evaluate how terrorists could obtain materials/expertise to produce a nuclear/radiological weapon. Although nuclear weapons could be stolen this is considered to be the most problematic option from the terrorist viewpoint, therefore, the 5 significant potential threats to international security are:

- Illegal trafficking of weapons grade nuclear material.
- Creation and deployment of dirty bombs.
- Sabotage and destruction of nuclear facilities.
- Sabotage of radiological materials in transit.
- Creation and deployment of nuclear weapons.

### 2.1: Trafficking of Weapons Grade Nuclear Material:

The trafficking of weapons grade nuclear material presents a serious threat to international security. Smuggling of drugs, precious stones and more recently human traffic has been increasing over recent decades. Nuclear smuggling is consistent with this trend and although attempts thus far have been in the main 'amateurish' it is likely that more competent smugglers will endeavour to acquire Highly Enriched Uranium (HEU) >20% or plutonium. (Woessner & Williams, 1996) [4]. Woessner & Williams (1996) assert;
*"None of the radioactive contraband that has been confiscated by Western authorities has been traced unequivocally to weapons stockpiles. Some of the plutonium that smugglers try to peddle comes from smoke detectors."* [4]. There is evidence that smugglers have attempted to deceive potential rogue buyers by offering mock fissile material.

---

[1] Former lecturer and post-doctoral researcher in the US who was responsible for a bombing campaign between 1978-1995. Central to his reasoning was the idea that modern society is enslaved to technology. This theme has deep philosophical roots in the beliefs of eco-activists throughout history. The Luddites in the early 19$^{th}$ century had similar misgivings about industrialisation and took to direct action.

The IAEA established an Illicit Trafficking Database (ITDB) in 1993 to record incidents of illicit trafficking. Eighty-nine (64.5%) IAEA member states participate in the database. As of December 31, 2005, IAEA listed 827 confirmed incidents involving the illicit trafficking in nuclear materials, including weapons-usable material. [5].

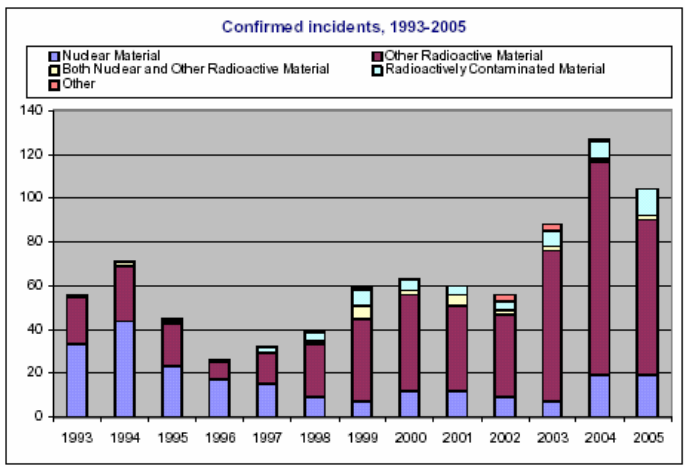

*Figure 1: Confirmed incidents of illicit trafficking in nuclear materials. Source: IAEA (2006) [5].*

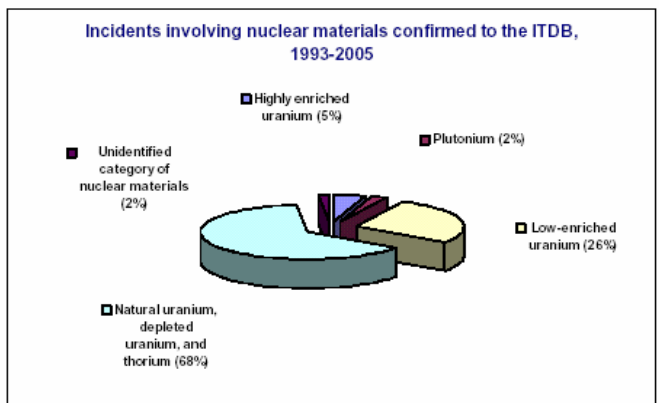

*Figure 2: Incidents involving different categories of nuclear materials confirmed to the ITDB 1993-2005. Source: IAEA (2006) [5].*

The US General Accounting Office (GAO) in 2003 stated that "20 involved either HEU (>20%) or plutonium-239". [6]. More than half of these incidents occurred during 1993-95. The remaining cases occurred during 1999-2001. In addition, IAEA (ITDB data) shows 1 further case of trafficking that involved 170g of HEU in Georgia (June 2003) [5].

This trend is of serious concern because the prospect of undetected weapons grade nuclear material being trafficked across borders is beyond accurate estimation.

It should be noted that IAEA member states that contribute to this database may not disclose all incidents. Illegitimate reasons for non-disclosure may include an unwillingness to share sensitive details that may undermine political confidence with implications for external relations. However, there may be legitimate reasons such as reluctance to publicise confidential material that might undermine current criminal investigations.

The US GAO (2003) state a number of key observations about 20 incidents involving weapons-grade nuclear material:
- Many of the incidents involved material that came from countries of the former Soviet Union, primarily Russia.
- Discovered nuclear material was seized primarily in Russia and Eastern Europe. There appears to have been an increase in trafficking in weapons grade nuclear material through the Caucasus (Georgia), Central Asia (Kyrgyzstan), Greece and Turkey. According to IAEA, it is uncertain whether the increase represents more trafficking in this material or better detection and reporting of activities that may have been going on in earlier years.
- Most of the smuggling incidents involved relatively small quantities of weapons-grade materials that were insufficient to construct a nuclear bomb. In some cases, the small quantities of material involved may indicate that the seller was trying to attract a potential buyer with a 'sample' quantity of material. In other cases, it appears doubtful that the traffickers had access to larger quantities of nuclear material.
- The incidents do not appear to be part of an organised criminal or terrorist activity or organisation[2].
- In most of the incidents, the weapons-grade material was seized as a result of police investigation. The material was not detected by

---

[2] Incidents seem to be too *ad hoc* to be associated with organised crime. However, this may be a view drawn from insufficient evidence and is therefore open to criticism.

equipment or personnel stationed by border crossings. One notable exception involved material detected by customs agents at a Bulgarian border crossing. This incident represents one of a few reported instances where nuclear material was shielded or protected to avoid detection. [6].

| Date | Details of interception |
|---|---|
| **May 1992** | Russia (Luch Scientific Production Association); 1.5 kilograms (90% HEU) was discovered by police investigation. |
| **May 1993** | Lithuania; 0.1 kilogram (50% HEU) was discovered by police investigation. |
| **July 1993** | Russia; 1.8 kilograms (36% HEU) discovered by police investigation. |
| **Nov 1993** | Russia; 4.5 kilograms (20% HEU) discovered by police investigation. |
| **March 1994** | Russia; 3.05 kilograms (90% HEU) discovered by police investigation. |
| **May 1994** | Germany; 0.006 kg of plutonium-239 discovered by police investigation. |
| **June 1994** | Germany; 0.0008 kg (87.8% HEU) discovered by police investigation. |
| **July 1994** | Germany; 0.00024 kg plutonium-239 discovered by police investigation. |
| **August 1994** | Germany; 0.4 kg of plutonium-239 discovered by police investigation. |
| **December 1994** | Czech Republic; 2.7 kg (87.7% HEU) discovered by police investigation. |
| **June 1995** | Czech Republic; 0.0004 grams (87.7% HEU) discovered by police investigation. |
| **June 1995** | Czech Republic; 0.017 kg (87.7% HEU) discovered by police investigation. |
| **June 1995** | Russia; 1.7 kg (21% HEU) discovered by police investigation. |
| **May 1999** | Bulgaria; 0.004 kg of HEU interdiction at border by Bulgarian customs. |
| **October 1999** | Kyrgyzstan; 0.0015 kg of plutonium-239 discovered by police investigation. |
| **April 2000** | Georgia; 0.9 kg (30% HEU) discovered by combination of radiation detection equipment at border and police investigation. |
| **September 2000** | Georgia; 0.0004 kg of plutonium-239 discovered by police investigation. |
| **December 2000** | Germany; <1 mg of plutonium-239 discovered in a forensic test. |
| **January 2001** | Greece; 0.003 kg of plutonium-239 discovered by police investigation. |
| **July 2001** | France; 0.005 kg (80% HEU) discovered by police investigation. |
| **June 2003** | Georgia; 0.17 kg HEU discovered by boarder police |

*Table 1: 21 cases of intercepted fissile material from trafficking (1992-2005).*

*Sources: US GAO (2003) & IAEA (2006) [5 & 6].*

The 21 incidents involving HEU (>20%) and plutonium-239 referred to above are detailed chronologically above.

Between 1992-2001 the US through a variety of agencies has spent over $86.1 million to enable over 30 countries (mostly the former Soviet Union and Eastern Europe) to defend against efforts to steal and smuggle fissile material. [6].

Radiation detection equipment such as hand-held detectors, stationary detectors at border crossings and airports, mobile vans with detection equipment and training have been provided. [6].

"While US assistance is generally helping countries combat the smuggling of nuclear and other radioactive materials, serious problems with installing, using and maintaining radiation detection equipment have undermined US efforts." (US GAO, 2003) [6].

Detection equipment provided for Lithuania was unnecessarily stored in the basement of the US embassy. Furthermore, it was found that few countries systematically report incepted contraband and therefore makes benchmarking and auditing of performance almost impossible.

It is estimated that the former Soviet Union amassed 30 000 nuclear weapons and 600 – 650 metric tons of weapons-grade material before it collapsed. The security of materials of key concern include:

- Caesium-137 (neutron reflector in reactors or bombs).
- Cobalt-60 (industrial and medical use but could be used in a 'dirty bomb').
- Lithium-6 (used in thermonuclear weapons).
- Plutonium-239 (used in nuclear weapons and reactor fuel).
- Uranium-235 (used in nuclear weapons and reactor fuel). [6].

It is acknowledged that the most likely smuggling routes are those established during the reign of the former Soviet Union to smuggle goods in, only in this instance it is to smuggle goods out. It is clear that ex-KGB and Soviet networks could be used for clandestine matters given Russia's impoverished workforce. Woessner and Williams (1996) suggest; "Some Turkish gangs appear to be engaged in the trade having graduated from clandestine export of antiquities, they treat uranium as just another commodity." [4].

It is understood that nuclear trafficking is in its embryonic stage but the potential for more sophisticated operations to develop grows with each passing year. Franchetti (1997) reported that in Russia; "The industry is seriously and dangerously under-funded – 70% of security devices at Russian facilities are outdated…Some of the staff working at these plants are desperately depressed and haven't been paid in months – and the temptation to smuggle nuclear material out is great. The situation is very serious." [7].

Difficulties arise where practical matters of implementing policies are inhibited by corruption. In 1998, 1500 Russian customs officials were dismissed for corruption. [1]. Moreover, the expanse of territory that borders Russia is immense and in some places treacherous. Border security is difficult and would be prone to opportunistic trafficking. Where borders lie in areas of hostile natural terrain, patrols and detection equipment act as a deterrent rather than as a solution.

Potter (1992) estimated that in the former Soviet Union there are between 1000-2000 individuals who have detailed knowledge concerning nuclear weapons design and between 3000-5000 who have experience of producing plutonium-239 and HEU. [8]. More recent estimates have not been found.

In November 1993, HEU was stolen from the Sevmorput shipyard near Murmansk, Russia. The material was recovered but the Investigation Officer, Mikhail Kulik asserted; "Even potatoes are probably much better guarded today than radioactive materials." [4].

The terrorist group Aum Shinrikyo who were responsible for the sarin gas attack on the Tokyo underground in 1995 have attempted to influence Russian scientists and research students through donating to leading facilities. It was further found that members of this cult had infiltrated the I. V. Kurchatov Institute of Atomic Energy and the Mendeleyev Chemical Institute. [1].

On the issue of 'rogue states' or sub-state actors recruiting Russian expertise, Cameron suggests; "…the majority of scientists that are moving are doing so to states that already possess a nuclear capability, such as Israel or China, and are intent on upgrading the quality of that capability." [1].

Furthermore, Aum Shinrikyo in 1993 attempted to meet with the Russian Minister of Nuclear Energy but this request was denied. After this rejection the cult attempted to arrange the purchase of natural uranium from Australia without success. [1].

It is for these reasons that the US has offered assistance to the states of the former Soviet Union through funding, equipment and expertise. At the core of this assistance is that it is better that fissile material is stored safely and accounted for *in situ* and that stolen fissile material is intercepted at national borders to prevent international trafficking. The US GAO (2003) refer to this as the 'first-line and second-line of defence' respectively and financial aid is directed towards meeting those two objectives. [6].

The prospect of a 'sub-state actor' moving components of a nuclear device to a target location and assembling a weapon *in situ* must not be overlooked. Detecting HEU is difficult if well shielded due to the low energy $\gamma$-rays emitted that cannot penetrate 2cm of lead and therefore escape the detection of scintillation monitoring equipment[3]. Plutonium-239 is more difficult to handle than HEU and has a pronounced signature, thus it would be more difficult to smuggle.

In addition, with regard to international action on potential nuclear terrorism threats the Pugwash Council (2003) reports; "…for too long there was too little concrete action by national governments and the international community to prevent such a catastrophe from occurring. In recent months, this has begun to change, most notably with the decision in June 2002 by the G8 countries to spend $20 billion over 10 years to eliminate large quantities of fissile material in Russia." [10].

This will involve the US pledging $10 billion and the remaining $10 billion matching this is to be pledged by the other G7 countries. Concerns must be raised as to whether this timescale (10 years) is too long. Historically the US has been pragmatic on this policy through pioneering the Co-operative Threat Reduction Programme (Nunn-Lugar agreement) initiated in November 1991. Its purpose is as follows:
- Fund states of the former Soviet Union to dismantle and destroy weapons of mass destruction.
- Strengthen the security of nuclear weapons and fissile materials in connection with dismantlement.
- Prevent proliferation and to enable demilitarisation of the industrial and scientific infrastructure that could support the production of weapons of mass destruction. [10].

Estimated global stocks of nuclear weapons and materials are displayed below:

---

[3] Problems arise at ports where scintillation monitoring equipment cannot distinguish between containers that hold plutonium-239 or containers that hold bananas that contain potassium-40. [9].

| Country | Total nuclear weapons[3] (including those in reserve) | HEU (metric tonnes)[4] Military[4] (1994) | Separated plutonium (metric tonnes) Military[4] (1994) | Separated plutonium (metric tonnes) Civilian[5] (2000) |
|---|---|---|---|---|
| US | ~9,000 | 580–710 | 85 | 0 |
| Russia | ~20,000 | 735–1365 | 100–165 | 34 |
| UK | <200 | 6–10[2] | 7.6 | 78.1 |
| France | ~350 | 20–30 | 3.5–6.5 | 82.7 |
| China | 410 | 15–25 | 2–6 | 0 |
| India | 30–35[1] | 0 | ~0.3 | 0 |
| Pakistan | 30–52[1] | 0.6–0.8 | 0.001–0.01 (end 1999) | 0 |
| Israel | 60–100 | 0 | ~0.4 | 0 |
| South Africa | 0 | 0.4 | 0 | 0 |
| North Korea | 0 | 0 | ~0.03 | 0 |
| Germany | 0 | 0 | 0 | 7.2 |
| Japan | 0 | 0 | 0 | 5.2 |
| Other European | 0 | 0 | 0 | 4.5 |
| Total | 30,085–30,152 | 1360–2140+ ~20 civilian | 200–270 | ~200 |

[1] Estimates based on the amount of nuclear material these states are believed to possess.
[2] 21.9 tonnes as published in the *Strategic Defence Review* 1998
[3] Carnegie Endowment for International Peace: http://www.ceip.org/files/nonprolif/numbers/default.asp
[4] Federation of American Scientists. *Public Interest Report* Vol.54, No. 6.
[5] Based on national declarations to the International Atomic Energy Agency (Infcircs549 http://www.iaea.org/worldatom/Documents/Infcircs).

Source: Nuclear Terrorism, Parliamentary Office of Science and Technology, Number 179, July 2002

*Table 2: Estimated global stocks of nuclear weapons and material.*
*Source: POST (2002) [11].*

There is huge scope for de-enrichment of Russian fissile material. An agreement between Russia and the US was reached in 1993. This complemented the Nunn–Lugar Co-operative Threat Reduction Agreement (1991) and required Russia to de-enrich 500 metric tons and sell this material to the US for civil reactors.

US commercial and political interests hampered the progress of this arrangement by increasing the timescale over a 20 year period, so as not to flood the international commercial uranium market resulting in the deflation of global uranium prices.

The Pugwash Council (2003) states; "….almost ten years after the original agreement, the material transferred to the US corresponds to less than 150 tons of Russian HEU (less than 30% of the target amount of 500 tons, and only 10-20% of all the HEU in the former Soviet Union)." [10].

**2.2: Dirty Bombs:**
The U.S Nuclear Regulatory Commission (2003) asserts; "A 'dirty bomb' or radiological dispersal device (RDD) is a conventional explosive or bomb containing radioactivity. The conventional bomb is used as a means to spread radioactive contamination….any type of material could be used in a dirty bomb, but in general this would be unlikely to cause serious health effects beyond those caused by the detonation of conventional explosives." [12].

The major threat of such a device is the impact of the explosion rather than the ensuing radiological contamination. Air dispersal would produce relatively low levels of radioactive contamination and consequently low doses to the unfortunates exposed. Clearly, those in the vicinity of such an exposure would attempt to move away from the epicentre of the explosion and in so doing reduce the likelihood of prolonged harmful exposure.

The detonation of a RDD in a contained environment (e.g. London Underground) could present greater numbers of casualties depending on the date and time of detonation. Emergency response teams would need to act quickly to minimise the radiation dose received by casualties. It should be noted that this scenario did not occur on the 7$^{th}$ July 2005 attacks in the London Underground but this does not mean that it could not happen in the future.

Contaminated areas would need to be cleaned up and this would be costly and time consuming. Dependent upon the time and site of the explosion it is difficult to estimate the potential economic disruption that would be caused by such an act.

Sources of radiological material that could be used in a 'dirty bomb' could derive from waste by-products from a nuclear reactor or medical waste. 'Orphan sources'[4] could be used that have become lost from regulatory control and management.

The Center for Defense Information (2001) raise concerns about Russia's security of nuclear waste. In 1996 Chechen rebels planted a 'dirty bomb' in Moscow's Izmailovo Park. The bomb was disengaged before detonation. It consisted of dynamite and caesium–137. [13].

**2.3: Sabotage of Nuclear Facilities:**
"…..nuclear facilities and materials….must be protected from mass-consequence sabotage. Securing these facilities and materials must be a top priority on the international agenda – something that must be pursued at every opportunity, at every level of authority, until the job is done". (Bunn & Bunn, 2002) [14].

Had one of the hijacked aircraft involved on the 11$^{th}$ September 2001 crashed into a nuclear power station, spent fuel pond, or reprocessing facility the results could have

---
[4] These are radioactive/nuclear materials that have bee abandoned, lost or stolen and are outside of regulatory control.

been catastrophic. If such an assault were carried out two potential scenarios emerge:
- Meltdown of the reactor core.
- Extensive dispersal of waste fuel on-site and potentially off-site.

Either scenario would result in extensive casualties and within the terms of analysis, nuclear sites act as *in situ* nuclear devices awaiting detonation. This scenario was envisioned by Ramberg (1984) in: *Nuclear Power Plants as Weapons for the Enemy: An Unrecognised Military Threat*. At the time Ramberg's view was dismissed as alarmist but times have changed. [15].

Nuclear power plants in most countries are designed with containment vessels several feet thick[5]. In addition, security includes armed guards and safety systems. Insider sabotage in co-operation with a well-armed terrorist cell would present a significant challenge to the most robust security systems. Moreover, the potential vulnerability of critical safety systems to truck bombs detonated outside the protected area of a power plant has been widely postulated in the literature.

Russia possesses nuclear power plants that are based on early Soviet design and do not have "….western-style containment vessels or the same level of redundant safety systems". (Bunn & Bunn, 2002) [14].

Research reactors are at risk from an attack but the consequences would be minor in comparison to a power reactor due to meagre inventories. There are research reactors located across the world and many contain weapons grade fissile material that is poorly secured.

Moreover, sabotage of a spent fuel pool resulting in a loss of cooling water could lead to a temperature in excess of $900^0$C. A zirconium fire ensuing could disperse radio-nuclides including; caesium-137, into the atmosphere.

A Design Based Threat (DBT) document (classified) is drawn-up by the Office for Civil Nuclear Security (OCNS) for all UK civil nuclear sites. The resilience of sites to attack is further advanced through the UK Nuclear Industries Security Regulations that provide for the strengthening of security regimes.

The application of 'strength in depth' is further manifest by the UKAEA Constabulary. On-site armed police protect Sellafield and Dounreay. This facility is not extended to other sites. [16].

---

[5] Nuclear power plants without containment structures include: RBMK (Soviet reactors), VVER-series (Soviet reactor) and MAGNOX (UK reactors). [16]

Edwards (2004) warns that in the UK "Evidence is emerging that the no-fly zones around nuclear plants are regularly breached by both military and civilian aircraft." No-fly zones around nuclear facilities have a radius 3.7km. "Over the last five years, the operators of 19 nuclear sites around Britain have lodged more than 100 complaints about aircraft flying too close." [17]. The Author considers that a no-fly zone of 3.7km is insufficient to enable a military response to an aircraft enclosing upon/targeting a nuclear power plant.

In the UK; "Measures [post 11th September 2001] have been taken to enable intervention by RAF interceptor aircraft in the event of an aircraft attack at a civil nuclear facility." [11].

France installed anti-aircraft missiles around the Cap de la Hague facility in immediate response to the events of 11th September 2001 but has since removed these defences. This is due to a perceived reduction in risk.

**2.4: Sabotage of Radiological Material in Transit:**
"Under UK regulations (based on IAEA recommendations) highly radioactive material is transported in robust flasks that would be difficult for terrorists to rupture." (POST, 2002) [11].

In the UK, the OCNS regulates the transportation of radiological material as arranged by civil nuclear authorities. The OCNS operates under the Convention on the Physical Protection of Nuclear Material for shipments sent and received.

Ships carrying radiological material (e.g. MOX fuel) have "...deck-mounted naval guns and an armed escort provided by the UKAEA Constabulary". [18]. Moreover, air, train and road transportation of radiological materials are packaged to withstand accidents according to the radiological hazard presented. Three categories are used:
- Type A – survives a minor impact.
- Type B – survives a major impact.
- Type C – survives an air accident.

There are no UK reports of deliberate sabotage of radiological material in transit. The increased potential threat of sabotage has been acknowledged throughout the global nuclear community and revision of the Convention on the Physical Protection of Nuclear Material was announced on 8th July 2005 by the IAEA [19]. Amendments include: improved protection measures for international transportation of nuclear materials and provides clarification on the definition of nuclear facilities and sabotage and appropriate levels of domestic protection.

**2.5: Development of Nuclear Weapons:**
**2.5.1: Fission Devices:**
Natural uranium contains approximately 98.28% uranium-238, 0.7% uranium-235 and 0.0058% uranium-234. Energy production is based on the fission of uranium-235, whereas uranium-238 must be transmuted through neutron capture to form the fissile product plutonium-239. Uranium-238 is regarded as a fertile material.

The prospect of rogue states or sub-state actors acquiring weapons grade fissile material (Highly Enriched Uranium – HEU >20% or plutonium) is of huge international concern. U-235 and Pu-239 in sufficient quantities can be used to produce weapons of mass destruction.

The Federation of American Scientists (2001) assert; "The minimum mass of fissile material that can sustain a nuclear chain reaction is called a critical mass and depends on the density, shape and type of fissile material, as well as the effectiveness of any surrounding material (called a reflector or tamper) at reflecting neutrons back into the fissioning mass." [20].

|  | Uranium-235 | Plutonium-239 |
| --- | --- | --- |
| Bare sphere | 56 kg | 11 kg |
| Thick tamper | 15 kg | 5 kg |

Table 3: Critical masses in spherical geometry for weapon-grade materials.
Source: FAS (2001) [20].

Uncertainty remains as to the minimum amount of plutonium-239 required to make a bomb. The implosion technique functions on the principle that the critical mass of compressed fissile material decreases as the inverse square of the density achieved.

Therefore, due to the decrease of the critical mass of the material as density of the material increases, it follows that 'Implosion' devices require limited amounts of fissile material compared to other types of fission weapons. Estimates between 8kg – 11kg have been espoused in the literature surveyed.
The implosion principle is illustrated below:

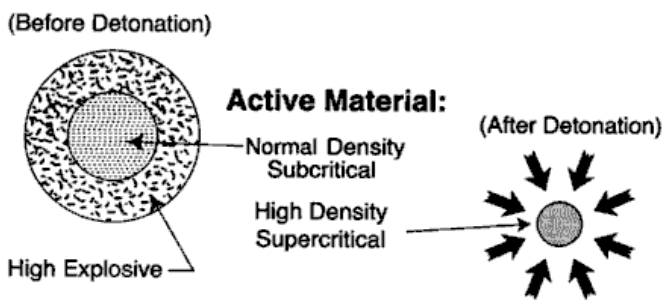

Fig.3: Implosion assembly principle.
Source: FAS (2001) [20].

From the point of view of construction the 'Gun' device would be the easiest to assemble. Two sub-critical masses are used in this device. One acts as a projectile and the other as a target. A propellant crashes these two sub-critical masses together to form a supercritical mass. Whereas plutonium-239 and HEU can be used in the 'implosion' design, only HEU can be used in the 'Gun' design.

The generic construction of a 'Gun' assembly is shown below:

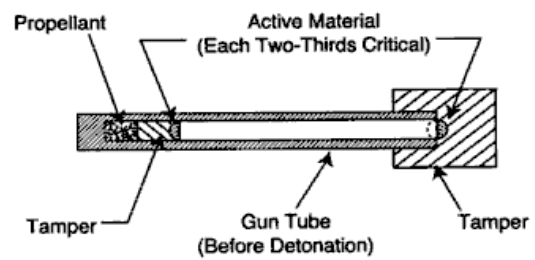

Fig.4: Gun assembly principle.
Source: FAS (2001)[20].

Techniques exist to enhance the radiation yield of a fission weapon and deuterium with tritium can be employed by means of fusion to generate increases in high-energy neutrons that encourage the fission process to increase its yield and decrease the overall size and weight of the weapon.

The Pugwash Council (2003) advanced; "HEU poses the danger that it is far easier to manufacture into nuclear weapons than is plutonium." [10]. Moreover, the open literature detailing how to construct a nuclear device is easily available, so that a technically competent team (e.g. non-state actor) with sufficient financial and technical resources could feasibly construct a nuclear weapon. (Allison, 2004) [21].

### 2.5.2: Fusion Devices:
Fusion (thermonuclear) devices utilise heavy isotopes of hydrogen (deuterium and tritium) to generate neutrons. When fused, intense heat and pressure is generated enabling the concomitant explosion to overcome the Coulomb barrier.

Thermonuclear weapons utilise the properties of lithium-6 deuteride ($^6$LiD, $\sigma_a$ = 70.5 barns at thermal energies – 2.2km/s) as a means of liberating tritium. Deuterium and tritium fuse creating alpha particles whose charge and high temperature assist in producing fire.

$$^6Li + n = {}^3H + {}^4He + Energy$$
(breaking down lithium to liberate tritium)
(Equation 1)

$$^2H + {}^3H = {}^4He + n + 17.6 MeV$$
(fusion reaction that causes ignition and liberates energy)
(Equation 2)

Lithium-6 is a controlled material and requires technical experience and knowledge due to its complex two-stage design. Moreover, lithium-6 is produced by the COLEX (column exchange) electrochemical process and few countries have the capability of making sufficient for large quantities of weapons. [20].

Deuterium ($^2H$) is necessary for the production of lithium-6 deuteride and is therefore subject to export controls. Heavy water ($D^2O$) is of relevance to issues of proliferation as it can be used in the production of weapons grade plutonium. Large-scale production of heavy water requires competent scientists and technical infrastructure because deuterium is 0.015% of the hydrogen in water and this percentage needs to be enriched to more than 99%. [20].

### 3.0: Preparedness and Emergency Response:
The IAEA dictum; prevent, detect and respond to nuclear threats has been clearly communicated to the international community and member states. Increased levels of preparedness are in evidence in the UK.

In the summer of 2004 the British Government sent-out the public information booklet: "Preparing for Emergencies: What You Need to Know". This was supported with television broadcasts. Advice was provided about how to prepare for an emergency such as a biological, chemical or radiological attack and helping to prevent a terrorist attack. MI5 produced further advice on their website. [22].

The British Government published the Civil Contingencies Act (2004) [23] to improve civil protection. In addition, legislation has been consolidated in response to the 11[th] September 2001 attacks. The Terrorism Act 2000 is extended through the Anti-Terrorism, Crime and Security Act 2001 [24].

Governments across the globe are confronting the prospect that their citizens may be subject to a terrorist attacks. Those attacks could involve radiological or nuclear weapons. Not all states will be able to respond with the same efficiency. Much depends upon the quality of the afflicted state's infrastructure and preparedness.

In the UK, contingency planning for nuclear accidents is well established. In the event of an accident detailed emergency response plans exist that extend to 15km from the nuclear facility. UK Parliamentary judgements will have to be made to decide whether emergency planning needs require an extension beyond that boundary.

Contingency planning for a RDD or nuclear device detonated in an urban area is in comparison a recent concern. The International Commission on Radiological Protection (ICRP) have drafted a report: "Protecting People Against Radiation Exposure in the Aftermath of a Radiological Attack" (2004). [25]. It is anticipated that this document will serve the authorities responsible for civil contingencies in every nation state at risk.

### 4.0: Outlook for Nuclear Security:
This paper has shown that the threat of terrorism involving nuclear or radiological weapons is credible. The technical knowledge to create nuclear or radiological weapons is available in the open literature. Eliminating unauthorised access to fissile or radioactive waste material (for the construction of a RDD) is the main safeguard.

However, terrorist organisations have attempted to obtain fissile material and trafficking remains a significant problem. Increased border control and international co-operation complemented with improved international and national legislative instruments will reduce the risk of an attack.

Nations with political and economic instability or theocratic governance are vulnerable to poor implementation of improved physical protection measures. Where workforce morale and corruption reside (such has been the case in the former USSR states) there is a danger that not enough will be done.

Opportunities for nuclear or radiological terrorism remain despite best efforts of protection. An attack on a nuclear power station or reprocessing facility is feasible despite current safeguards in the UK and worldwide.